# RAFDA: A Policy-Aware Middleware Supporting the Flexible Separation of Application Logic from Distribution

Scott M. Walker, Alan Dearle, Stuart J. Norcross, Graham N. C. Kirby & Andrew J. McCarthy
School of Computer Science, University of St Andrews, St Andrews, Fife, Scotland.
{scott, al, stuart, graham, ajm}@cs.st-and.ac.uk

## ABSTRACT
Middleware technologies often limit the way in which object classes may be used in distributed applications due to the fixed distribution policies that they impose. These policies permeate applications developed using existing middleware systems and force an unnatural encoding of application level semantics. For example, the application programmer has no direct control over inter-address-space parameter passing semantics. Semantics are fixed by the distribution topology of the application, which is dictated early in the design cycle. This creates applications that are brittle with respect to changes in distribution.

This paper explores technology that provides control over the extent to which inter-address-space communication is exposed to programmers, in order to aid the creation, maintenance and evolution of distributed applications. The described system permits arbitrary objects in an application to be dynamically exposed for remote access, allowing applications to be written without concern for distribution. Programmers can conceal or expose the distributed nature of applications as required, permitting object placement and distribution boundaries to be decided late in the design cycle and even dynamically. Inter-address-space parameter passing semantics may also be decided independently of object implementation and at varying times in the design cycle, again possibly as late as run-time. Furthermore, transmission policy may be defined on a per-class, per-method or per-parameter basis, maximizing plasticity. This flexibility is of utility in the development of new distributed applications, and the creation of management and monitoring infrastructures for existing applications.

## Keywords
Middleware, Java, Distributed Systems, POJO.

## 1. INTRODUCTION
Existing middleware systems suffer from several limitations which restrict the kinds of application that can be created using them and hamper their flexibility with respect to distribution and adaptability. In this paper we focus on four of these limitations, namely,

1. They force decisions to be made early in the design process about which classes of object may participate in inter-address-space communication.
2. They are brittle with respect to changes in the way in which the applications are distributed.
3. It is difficult to understand and maintain distributed applications since the use of middleware systems may force an unnatural encoding of application level semantics.
4. It is difficult to control the policy used to determine how objects are transmitted among the available address-spaces in a distributed application.

**Early Design Decisions** – Existing middleware systems all require the programmer to decide at application design time which classes will support remote access and to follow similar steps in order to create the remotely accessible classes. The programmer must decide the interfaces between distribution boundaries statically then determine which classes will implement these interfaces and thus be remotely accessible. These classes, known as *remote classes*, are hard-coded at the source level to support remote accessibility and only instances of these classes can be accessed from another address-space. Therefore, the programmer must know how the application objects will be distributed at run-time before creating any classes.

Some middleware systems require the manual creation of ancillary code such as skeletons, proxies and stub implementation classes, which must extend special classes, implement special interfaces or handle distribution related error conditions, based on programmer-defined interfaces. All require the creation of server applications that configure the middleware infrastructure then instantiate and register objects for remote access.

**Brittleness with Respect to Change -** A distributed application created using an existing middleware system is brittle with respect to change because the distribution of the application must be known early in the design process. The possible partitions of a distributed application are dependent on which classes within the application support remote access, restricting the classes of object that can be referenced across address-space boundaries.

**Distorted Application Level Semantics** - Existing middleware systems force remotely accessible classes to extend special classes, implement special interfaces or handle network related errors explicitly. It is not possible to make application classes remotely accessible unless their super-classes also meet the necessary requirements. At best, this forces an unnatural or inappropriate encoding of the application semantics because classes are forced to be remotely accessible for the benefit of their sub-classes and, at worst, application classes that extend library classes cannot be remotely accessible at all.

**Inflexible Parameter Passing Semantics** - Existing middleware systems decide the parameter passing semantics applied during remote method call statically based on the remote accessibility of the application classes. The parameter passing semantics is tightly bound to the distribution of the application so changes to the distribution of an application have the side effect that application semantics may be altered. All objects of the same class must be transmitted in the same way, whether this is appropriate or not, and the programmer does not have the freedom to choose different parameter passing semantics for classes on a per-application or per-call basis.

## 2. RELATED WORK

The creation of remotely accessible objects using industry standard middleware systems such as Java RMI [1] and Microsoft .NET Remoting [2] requires the programmer to take the following steps:

- The programmer is forced to decide statically the interfaces between distribution boundaries.
- The programmer is forced to decide statically which classes of component will implement these interfaces and thus be remotely accessible.
- These remotely accessible classes must extend a special base class that provides the functionality necessary for remote accessibility. This has two effects: to force the static identification of accessible classes, as above, and, in languages without multiple inheritance, to prevent the creation of accessible subclasses of existing non-accessible classes.
- Once a remotely accessible class is instantiated, the instance is associated with a naming service that allows remote callers to obtain a remote reference to it.

Only instances of classes that support remote access may be separated into different address-spaces from their reference holders, constraining the ways in which applications can be distributed. To change an application's distribution, programmers may be forced to introduce support for distribution into classes without it. Conversely, programmers must determine whether the additional application complexity inherent in unnecessarily supporting remote access outweighs the cost of removing it in terms of programmer effort.

Without the ability to expose objects to remote access dynamically, application distribution is inflexible. Yet, it is not possible to introduce support for remote access into every application class using existing systems because of the semantic restrictions placed on remote classes. For instance, application classes cannot pass remote references to instances of pre-defined library classes that do not support remote access.

Thus, the requirement to follow the above steps leads to the problems of inflexibility exhibited by industry standard middleware systems with respect both to static design-level changes and to dynamic run-time changes in application partition. Programmers can attempt to overcome these problems by explicitly creating distributed objects that can access objects for which static type information was not available at compilation time. The *tie* approach and *Dynamic Skeleton Interface (DSI)* provided by CORBA [3] illustrate the manner in which this can be achieved.

Using the CORBA DSI, specially constructed remotely accessible objects can extract operation names and any associated arguments from incoming remote method call requests, then perform the requisite calls. CORBA does not serialize objects in a self-describing manner and so programmers must write code to extract type information from the requests and deserialize the arguments. Consequently, it is possible to construct applications in which methods are invoked with inappropriate arguments [4]. This leads to unexpected application semantics and may cause run-time problems in strongly typed languages.

CORBA DSI and similar ad-hoc approaches to increased flexibility can be adopted only at the cost of increased complexity. Programmers lose the abstraction over the inter-address-space communication afforded when using proxies and objects created from IDL. Middleware level functionality must be implemented at the application level, obviating the primary benefits that middleware systems offer.

Several research-based systems have been developed in order to overcome the limitations of industry standard systems. However, these systems only partially tackle the limitations of industry standard systems.

#### *2.1.1 Early design decisions*

Some research based systems, such as JavaParty [5] and Do! [6], employ custom compilers to generate distribution-related code based on source code that has been annotated by the programmer. Consequently, programmers are still forced to make early design decisions. These systems simplify the process of creating distributed applications through automated code generation but it is the programmers that must determine which classes will support remote access.

Tools such as J-Orchestra [7] and Pangaea [8] are designed to transform a single non-distributed application into an isomorphic version that distributes itself across the distributed system at run-time. These systems perform static code analysis to help programmers choose suitable distributions. The distributed version of an application is generated automatically and in this respect these systems allow separation of application logic from distribution. However, both transform only local applications and are unsuitable tools for the creation of general distributed applications since programmers cannot include multiple entry points. Programmers define initial application partitions using the provided tools and though both systems support changes to application distribution using migration, it is not possible to migrate arbitrary application objects. If fundamental changes are made to application distributions then the applications must be re-transformed, limiting the effectiveness of these systems in dynamically changing systems.

ProActive [9] and JavaSymphony [10] allow programmers to expose objects to remote access dynamically. However, both adopt the active object [11] model which associates a thread with each remotely accessible application object. The conversion of existing application objects into active objects alters the threading semantics of the application. Further, active objects may not have shared access to any non-active objects. Programmers may need to alter the structure of the application to ensure that this strict separation of active object closures is preserved.

The *JBoss Enterprise Middleware System* [12] provides *AOP Remoting*, which uses aspect-oriented programming techniques to instrument instances of existing classes for remote access. AOP Remoting allows the exposure of application objects to remote access as services using SOAP [13] or Java RMI. AOP Remoting places semantic restrictions on the classes of object that can be exposed. All classes must provide default constructors and all method arguments and return values must be *Serializable*. AOP Remoting adopts a service-oriented model, rather than a complete Distributed Object Model, in which methods of the underlying objects are provided to remote clients, providing the objects meet the above semantic requirements.

### *2.1.2 Brittleness with respect to change*

In addition to forcing decisions early in the design process, which results in static inflexibility to change, distributed applications created using existing middleware systems also exhibit brittleness with respect to dynamic change. Brittleness and inflexibility to change occurs in middleware systems that do not allow dynamic re-partitioning of applications, hindering the adaptability of those applications to changing execution environments. For example, objects on heavily loaded machines cannot migrate to other machines. It also has implications for long running systems as applications cannot be re-distributed as machines join and leave the distributed system. Several of the research based systems support object mobility, including JavaParty [5], J-Orchestra [7], ProActive [9], JavaSymphony [14] and Pangaea [8].

### *2.1.3 Distorted application level semantics*

Application semantics are affected by the restrictions placed on application classes that support distribution. Inheritance relationships between classes are affected and it is difficult to make application classes remotely accessible if their super-classes do not meet the necessary requirements. This causes an unnatural or inappropriate encoding of application semantics because classes are forced to support remote access for the benefit of their sub-classes, entangling application logic and distribution. This is particularly a problem for application classes that need to extend pre-compiled classes without support for remote access.

Industry standard middleware systems decide statically which parameter passing semantics should be applied when remote methods are called. In Java RMI [15], only classes that implement the *java.rmi.Remote* interface and handle network related errors explicitly in application logic can be exposed to remote access or passed by-reference. All other objects that are passed as arguments or return values to remote methods must be instances of classes that implement the *Serializable* interface. Parameter passing semantics are affected by static design level decisions and are tightly coupled with application distribution.

Microsoft .NET remoting [16] adopts semantics that are similar to Java RMI. Instances of classes that extend the *MarshalByRefObject* class are passed by-reference and all other objects that are passed to remote methods must be instances of *Serializable* classes. The .NET remoting framework incrementally improves on Java RMI by applying these semantics consistently to objects. However, parameter passing semantics are still fixed statically and are dependent on the distribution of the application.

In CORBA and COM, arguments are marked in IDL with the passing semantics to be applied. Further, CORBA component classes are defined statically as either pass-by-reference or pass-by-value. CORBA and COM allow only components and data structures that have been explicitly described to be passed across address-space boundaries.

The research-based middleware systems strive to preserve local Java method calling semantics and so fix parameter passing semantics statically. Consequently, programmers cannot employ the most advantageous parameter passing semantics for the circumstances of each application. Programmers cannot take control over application semantics, hindering the reuse of library classes in distributed contexts since the parameter passing semantics cannot be specified independently of class implementation.

In general, reusability and application semantics are restricted for the following reasons:

**Some systems allow no programmer control over parameter passing semantics at all** - These systems lack flexibility as programmers cannot employ the most suitable parameter passing mechanisms on a per-application basis. With control over passing semantics, programmers can manage the trade-offs between different parameter passing mechanisms to reduce network traffic, introduce resiliency or permit caching.

**When programmers can decide parameter passing semantics, they cannot do so dynamically** - Application programmers have limited dynamic control over inter-address-space parameter passing semantics. Within a single application, it may be required that objects are transmitted by-value or by-reference depending on the circumstances; in most existing middleware systems this would require that different classes be created.

Complexity is introduced into applications due to the limitations of the middleware system.

**The parameter passing semantics and application distribution are tightly bound** - The parameter passing semantics and application distribution are tightly coupled. Reuse of large-grained components, composed of instances of multiple classes, is hindered because concrete class implementations must be developed in the context of some planned deployment environment. Various physical considerations dictate the nature of the implementation, such as the available computational resources, network connectivity, latency or bandwidth. These considerations influence the implementation of classes, limiting reuse [17]. For example, in a poorly connected environment, it may be appropriate that pass-by-value semantics are adopted in order that the called methods can continue to perform computation over arguments, even if the network connection to the caller is lost transiently. Conversely, in a well-connected environment, it may be appropriate to adopt pass-by-reference semantics to allow shared access to arguments and ensure coherency.

## 3. THE RAFDA SYSTEM

This paper introduces RAFDA [18, 19] a Java middleware system that provides control over the extent to which inter-address-space communication is exposed to programmers, in order to aid the creation, maintenance and evolution of distributed applications. The described technology adopts a *plain old Java object (POJO)* approach and permits arbitrary application objects to be exposed for remote access dynamically. Object instances are exposed as Web Services [20] through which remote method invocations may be made. RAFDA has four notable features that differentiate it from other middleware technologies.

1. The programmer does not need to decide statically which application classes support remote access. Any object instance from any application, including compiled classes and library classes, can be exposed as a Web Service without the need to access or alter the application's source code.
2. The system integrates the notions of Web Services and Distributed Object Models by providing a remote reference scheme, synergistic with standard Web Services infrastructure, extending the pass-by-value semantics provided by Web Services with pass-by-reference semantics. Specific object instances rather than classes are exposed as Web Services, further integrating the Web Service and Distributed Object Models. This contrasts with systems such as Apache Axis [21] in which only classes are exposed as Web Services.
3. Parameter passing mechanisms are flexible and may be dynamically controlled through policies. An exposed component can be called using either pass-by-reference or pass-by-value semantics on a per-call basis.
4. The system automatically exposes referenced objects on demand. Thus an object *b* that is returned by method *m* of exposed object *a* is automatically exposed before method *m* returns.

The process of implementing the application logic is thus separated from the process of distributing the application. Since any object can be made remotely accessible, changes to distribution boundaries do not require re-engineering of the application, making it easier to change the application's distribution topology. This separation of concerns simplifies the software engineering process to the programmer's advantage, both when creating a distributed application and introducing distribution into an existing application. This simplifies the creation of tools such as monitoring and management components that need to access and modify object state from outwith those objects' local address space. Using traditional middleware systems, it is difficult to attach such tools to existing objects without access to source code and extensive engineering effort.

This functionality is provided by the RAFDA Run-Time (RRT), a middleware system for Java development that tackles the problems inherent in existing middleware systems. The RRT simplifies the kinds of tasks that are common to the creation of distributed application such as dynamically exposing objects for remote access, obtaining remote references to remotely accessible objects, and remote method invocation.

The RRT conceals the complexity of distribution where appropriate, allowing distribution to be introduced into applications quickly. This reduces the software engineering effort required to create distributed applications, leading to quick application prototyping. However, the RRT also permits programmers to expose aspects of application distribution as required, allowing the creation of applications that can exploit their distributed nature and are flexible with respect to change. The RRT has advantages over traditional middleware approaches as it adapts its behaviour to suit the requirements of a given distributed application, rather than forcing the programmer to adapt the application to the requirements of the middleware system.

Applications access the functionality provided by the RAFDA system by calling methods on infrastructure objects called *RRTs*. There is an RRT in each address-space in the distributed system, analogous to a CORBA ORB. Each RRT provides two interfaces to application programmers. The first, called *IRafdaRunTime*, provides server-side operations to application objects collocated with the RRT, allowing programmers to expose objects or access frameworks that control transmission policy and distribution policy. The second, called *IRafdaRunTime-Remote*, provides client-side functionality to application objects that are remote with respect to the RRT, allowing programmers to obtain remote references to existing objects or to perform object migration.

Figure 1 shows the RRT instances present in two address-spaces. The large circles represent objects in the distributed application. Each RRT instance is represented by a shaded box with the *IRafdaRunTime* and *IRafdaRunTimeRemote* interfaces shown. Each RRT is accessible locally via the *IRafdaRunTime* interface and remotely via the *IRafdaRunTimeRemote* interface.

In addition to the functionality examined in this paper, the RRT provides remote object instantiation, object migration and a distribution policy framework that is used to automate object placement. A complete description of the RRT and its implementation, including the features described in this paper, can be obtained in Walker [22]. Although the RRT is written in Java and is designed to support Java, it does not employ any language-specific features unique to Java. The techniques described here are applicable in other languages.

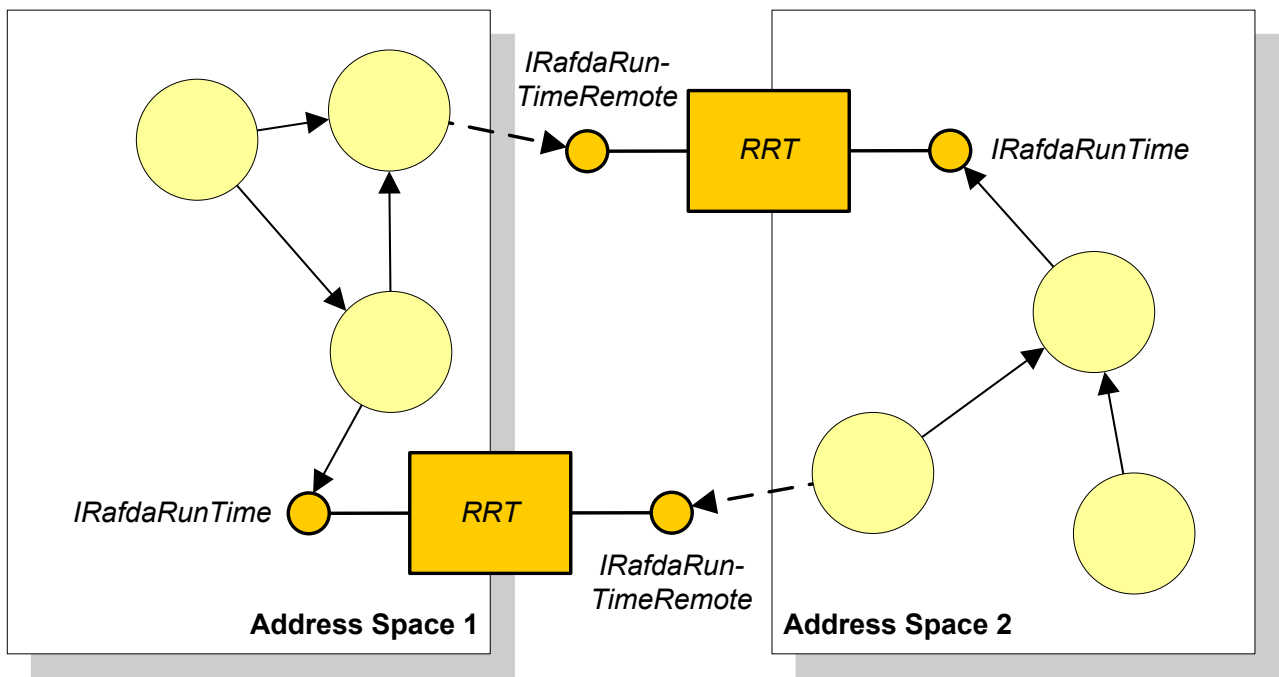

**Figure 1: RRT instances exposing different interfaces to local and remote objects.**

## 4. EXPOSING ARBITRARY OBJECTS FOR REMOTE ACCESS

The RRT permits arbitrary application objects to be exposed for remote access. Specific application objects, rather than application classes, are exposed via Web Services. In order to make an object remotely accessible it is first registered with the RRT. Registration of an application object creates a Web Service running within the RRT that uses the exposed object as the underlying *service object* on which incoming Web Service requests are performed. In effect, the RRT maps Web Service requests to method calls on object instances and performs appropriate encoding of the results. Exposed objects may be referenced by other local objects; neither the reference holders nor the exposed objects are aware that registration has taken place.

Each RRT implements the *IRafdaRunTime* interface shown in Figure 2. Only a subset of the functionality provided by this interface is shown. The omitted methods are used to control object migration and to automate object placement using programmer-defined policies.

```
public interface IRafdaRunTime {
  void expose(Object objectToExpose,
    Class remoteType, String serviceName);
  TransmissionPolicyManager
    getTransmissionPolicyManager();
  void associateClassWithRemoteType(
    Class applicationClass,
    Class remoteType);
  /* Other methods omitted */
}
```

**Figure 2: A subset of the *IRafdaRunTime* interface.**

This interface provides the *expose()* method, used to expose an object to remote access, the *getTransmissionPolicy-Manager()* method, used to control the transmission policy defining parameter passing semantics, and the *associateClassWithRemoteType()* method, used to control automatic exposure. The latter two methods are examined in Sections 6.1 and 7.4 respectively.

The *expose()* method takes three parameters to specify the object to be exposed, a *remote type* (that is, the interface that the exposed object should provide to remote clients) and a logical name for accessing the object. A number of issues arise from this simple method. Firstly, the *objectToExpose* may be any 'Plain Old Java Object' (POJO), so need not implement any special interfaces or extend any particular classes, maximizing flexibility. Secondly, the *objectToExpose* need not implement the interface specified in the *remote type* parameter although it must be structurally compliant with that interface. This again maximizes flexibility and permits classes to be exposed to remote access even if they were not envisioned to be so at design time. The *remote type* parameter can be a class or an interface; in either case, the method signatures are extracted to form the Web Service interface for the exposed object. The *remote type* parameter is optional. If omitted, the object is exposed with an interface matching its concrete type.

The *remote type* is the distributed equivalent of an interface in a non-distributed application class and is used to control method visibility. It is supplied on a per-object, not a per-class, basis. Any method can be made remotely accessible, irrespective of its local protection modifier. By default, RAFDA will preserve local protection semantics in the distributed application when the RRT is used both client and server-side, but allows only *public* methods to be invoked when using standard Web Services technology in order to preserve code encapsulation.

Remote types provide multiple views over exposed objects to remote clients. From the perspective of clients, exposed objects are instances of their associated remote types. Different instances of a single class can be exposed with different remote types and a single object can be exposed multiple times with different remote types. This allows the programmer to expose a single object with different logical names and different interfaces. Exposure can fail, resulting in a runtime exception, if the *remote type* contains methods that do not exist in the class of the object being exposed.

The *serviceName* parameter permits the exposed object to be addressed using a logical name which must, of course, be unique within the local address space.

An object of any class can be exposed, including precompiled classes and those with native members. There are two caveats. Firstly, the Web Services model provides no facility to allow field access, only method call. Thus the fields of an exposed object cannot be directly accessed and if the object does not provide *get()* and *set()* accessor methods then the fields cannot be accessed at all. This is a problem for all Java middleware systems since field access cannot be intercepted. Secondly, the current RRT implementation does not permit *remote types* to be final classes or to contain final methods. Exposure will fail if attempted using such a remote type. Note that no restrictions are placed on the classes of object that can be exposed, only on the remote types that may be applied to those objects. The RRT provides a class loader that can be used to change application classes and methods such that they are non-final to overcome this limitation. However, the class loader cannot transform system classes dynamically, meaning that system classes that are final or contain final methods cannot be used as remote types.

To illustrate the use of *expose()*, we use a small Peer-to-Peer (P2P) application as an example. A programmer has implemented a class called *P2PNode* which represents a node in a P2P routing network. This class is shown in Figure 3. This class has not been written with concern for distribution and does not implement any special interfaces or extend any base classes.

```
public class P2PNode {
  private final Key key;
  public P2PNode(Key key){…}
  public void addPeer(P2PNode peer){…}
  public void route(Key key, Message
msg){…}
  public String getLog(){…}
  public void stop(){…}
  public void start(){…}
  public Key getKey();
}
```

**Figure 3: The *P2PNode* implementation.**

The programmer obtains a reference to the *IRafdaRunTime* interface provided by the local RRT using the static method *RRT.get()*. Figure 4 shows how another programmer could expose an instance of this class as part of some P2P application. The programmer wishes to expose the functionality of the node using three different interfaces — a management interface for controlling the node remotely, a monitoring interface and an interface exposing the P2P functionality. These interfaces are named *IManage*, *IMonitor* and *IP2PNode* respectively. Each of these interfaces is associated with the names *Manage*, *Monitor* and *P2P* respectively. It is assumed that these are well known names that are used by client programmers to access the services.

```
public interface IManage {
  void stop();
  void start();
}
public interface IMonitor {
  String getLog();
}
public interface IP2PNode {
  public void addPeer(P2PNode peer);
  public void route(Key key, Message msg);
  public Key getKey();
}
public class ExposeP2PNode {
  public static void main(String[] args) {
    P2PNode p2pNode = new P2PNode(
      new Key());
    IRafdaRunTime rrt = RRT.get();
    rrt.expose(p2pNode, IManage.class,
      "Manage");
    rrt.expose(p2pNode, IMonitor.class,
      "Monitor");
    rrt.expose(p2pNode, IP2PNode.class,
      "P2P");
  }
}
```

**Figure 4: Exposing an instance of class *P2PNode*.**

## 5. CLIENT-SIDE DISTRIBUTED OBJECT PROGRAMMING USING THE RRT

Exposed objects may be accessed either using their service names or Globally Unique Identifiers (GUIDs) allocated to the associated services at exposure time. Both of these may be discovered dynamically by clients. Typically, an application will expose a small collection of objects with well known names thus avoiding the need for dynamic GUID discovery. Exposed objects may be addressed using URLs of the following form:

```
http://<host>:<port>/<serviceName|GUID>
e.g. http://host.rafda.org:5001/P2P
```

As stated previously, the RRT implements an interface called *IRafdaRunTimeRemote*. A subset of this interface, through which client-side programmers access remotely accessible objects, is shown in Figure 5. The omitted methods are used to perform remote instantiation of objects, to migrate objects between address-spaces and to control automated object distribution based on programmer-defined policies.

```
public interface IRafdaRunTimeRemote {
  Object getRemoteReference(
    String serviceName);
  /* Other methods omitted */
}
```

**Figure 5: A subset of the *IRafdaRunTimeRemote* interface.**

The *IRafdaRunTimeRemote* interface provided by an RRT contains a method called *getRemoteReference()* that permits a handle to be obtained to any object exposed by that RRT. As will be shown later, the handle returned may be a reference to a proxy for a remote object, a local copy of the object or a hybrid of the two (a smart proxy). The *getRemoteReference()* method takes an argument that identifies the service name with which the requisite object was exposed. The name can be either the programmer-defined service name or the automatically generated object GUID.

The object returned by *getRemoteReference()* can be cast to the remote type of the exposed object. Figure 6 shows the client-side code necessary to use the *P2PNode* exposed in Figure 4. The object returned by *getRemoteReference()* is cast to type *IP2PNode* which was the interface used as its remote type.

Programmers can obtain a remote reference to the *IRafdaRunTimeRemote* interface provided by a remote

RRT based on the socket address to which that RRT is bound using the static *RRT.getRemote()* method.

```
public class P2PClient {
  InetSocketAddress isa = new
  InetSocketAddress("host.rafda.org",
5001);
  IRafdaRunTimeRemote remoteRRT =
  RRT.getRemote(isa);

  public void deliver(Key dest, Message
msg)
  throws Exception {
    IP2PNode node = (IP2PNode)
    remoteRRT.getRemoteReference("P2P");
    node.route(dest, msg);
  }
}
```

**Figure 6: Client side code accessing a remote *P2PNode*.**

## 5.1 Browsing Exposed Objects

As described, distributed applications are bootstrapped by accessing objects based on their service names. The RRT provides a web interface that can be accessed using a conventional web browser to obtain human-readable information about exposed objects. Each exposed service is listed, showing the remote type, the URL, the real class of the exposed object and a string representation of the service object.

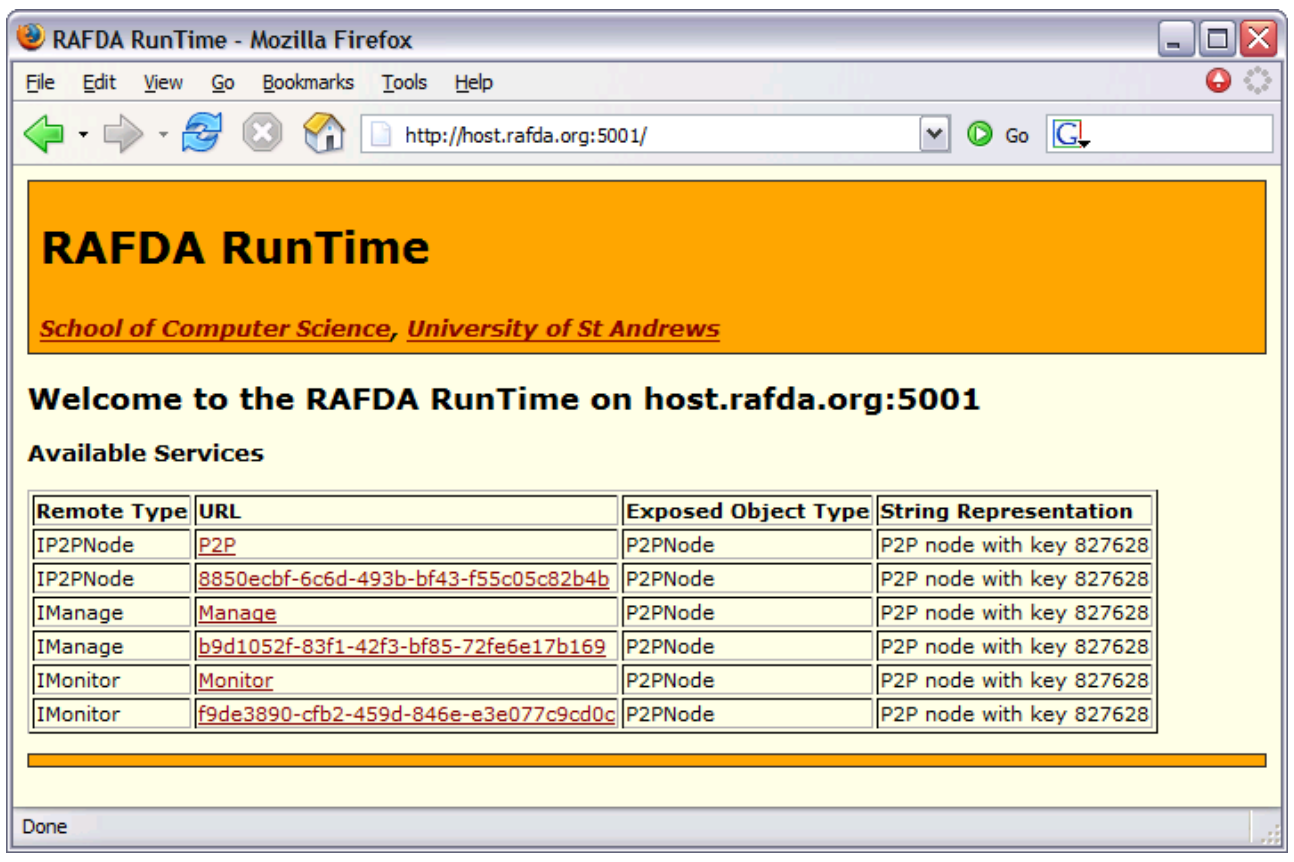

**Figure 7. Browsing an RRT.**

The links in the URL column refer to service-specific pages that provide:

- A list of the methods provided by the remote type.
- A list of the methods and fields provided by the exposed object's class.
- The current state of these fields in the exposed object.

By default, RRT instances show information only about the remote types. The information about the underlying exposed object is not available unless this functionality is explicitly enabled in the RRT configuration.

## 5.2 Failure

Distributing an application introduces new failure modes. The RRT treats network failure differently from application failure. Application exceptions are always thrown back to clients as they are not the concern of the RRT. Distribution-related exceptions are either handled directly by the RRT or propagated back to clients according to the RRT configuration.

Distribution-related exceptions are wrapped in *unchecked exceptions*. In Java, methods do not need to declare statically that they throw unchecked exceptions and callers are not forced to define handlers. Thus, there are three approaches to handling distribution-related errors that are open to developers:

1. Configure the RRT to handle all distribution-related exceptions internally. If failure occurs, default values (null, zero, etc.) are returned. No application level distribution-related exception handlers need to be defined in this case.
2. Configure the RRT to propagate all distribution-related exceptions to the clients but do not define application level exception handlers. If failure occurs, the uncaught exception causes the RRT instance to terminate immediately.
3. Configure the RRT to propagate all distribution-related exceptions to the clients and define application level exception handlers statically at any points in the application where failure can occur. If a distribution-related exception occurs, it is handled in a programmer-defined manner.

Systems such as Java RMI require programmers to handle any potential distribution-related errors explicitly at any points where remote calls are performed. In contrast, by providing a multiplicity of approaches to handling failure, the RRT simplifies application prototyping as programmers can ignore the possibility of distribution-related exceptions during initial development. The RRT offers programmers the flexibility to introduce error handling code into applications only where it is deemed necessary.

# 6. CONTROLLING OBJECT TRANSMISSION POLICY

As described in the introduction, using traditional middleware, the distribution topology of an application determines the object transmission semantics that are employed during remote method calls. For example, in Java RMI [1], only classes that implement the *java.rmi.Remote* interface and meet certain other criteria may be exposed for remote access. Such objects are always passed by-reference if they are accessed across an address space boundary. All other objects that traverse address-space boundaries must be instances of classes that implement the *java.io.Serializable* interface and these objects are always passed by-value. Similar problems can also be observed in Microsoft .NET Remoting [2], CORBA [3] and Web Services [20].

Within a single application, it may be required that instances of some class are transmitted by-value or by-reference depending on the circumstances. In most existing middleware systems this would require that different classes be created. Further, hybridisation is sometimes desirable, whereby some object state is cached at a client whilst other state is remotely accessed. Using the RRT's transmission policy framework, the application programmer can employ the most advantageous object transmission policy for the circumstances.

In addition to providing the programmer with the flexibility to control the application semantics, the dynamic specification of policy independently of class implementation allows the roles of library class programmer and application programmer to be separated. The library class programmer is concerned only with the functional requirements. Thus, library classes make fewer assumptions about the environment in which they are to be exposed. The application programmer has the freedom to apply any parameter passing policy to instances of any class, increasing the likelihood that any given class will be reusable in another context.

## 6.1 Defining Transmission Policy

By default the RRT passes objects by-reference when interacting with other RRTs and by-value when interacting with standard Web Service clients. However, the transmission policy framework described here provides a mechanism to allow the programmer to specify dynamically how objects should be transmitted during inter-RRT remote method calls. This is achieved using the local RRT's *transmission policy manager*, the interface to which is shown in Figure 8. This *TransmissionPolicyManager* interface provides methods through which the six different types of supported policy rule can be set. Methods to evaluate the currently active transmission policy are omitted. Programmers obtain a reference to the *TransmissionPolicyManager* interface using the *getTransmissionPolicyManager()* method provided by *IRafdaRunTime*.

```
public interface TransmissionPolicyManager
{
  /* Setting transmission policies */
  void setMethodPolicy(
    Method methodIdentifier,
    PassingMechanism passingMechanism,
    int depth, int priority);
  void setReturnPolicy(
    Method methodIdentifier,
    PassingMechanism passingMechanism,
    int depth, int priority);
  void setArgumentPolicy(
    Method methodIdentifier,
    int argumentNumber,
    PassingMechanism passingMechanism,
    int depth, int priority);
  void setClassPolicy(
    Class classIdentifier,
    PassingMechanism passingMechanism,
    int priority);

  /* Caching */
  void setFieldToCache(
    Field fieldIdentifier,
    Method getMethodIdentifier,
    Method setMethodIdentifier);
  void setMethodToCache(
    Method methodIdentifier);

  /* Other methods omitted */
}
```

**Figure 8: The *TransmissionPolicyManager* interface.**

The six types of rule supported by the transmission policy framework are as follows:

- *Method policy rules* are associated with methods as a whole and are set using the *setMethodPolicy()* method. This method specifies how method arguments should be transmitted. For example, a method policy rule might specify that during a call to a particular method, the arguments should all be passed by-reference. The parameters to *setMethodPolicy()* include the identity of the method to which the policy applies, the policy to be applied (using constant values of the enumeration type *PassingMechanism*, which is not shown here), the depth to which the closure of the parameters should be traversed in the case of pass-by-value, and a rule priority (discussed below).
- *Return policy rules*, set using the *setReturnPolicy()* method, are also associated with methods but control how the return values from methods should be transmitted. For example, a return policy rule might specify that the return value from a particular method should be passed by-value. The method policy rule and return policy rule associated with a single method are independent of each other and need not specify the same behaviour. The *setReturnPolicy()* method takes the same arguments as the *setMethodPolicy()* method which apply to the return value rather than the parameters.
- *Argument policy rules*, set using the *setArgumentPolicy()* method, are associated with individual method arguments and indicate how particular arguments within a method signature should be transmitted. They allow the programmer fine-grained control over the policy that is applied to each of the arguments of a method. The parameters to this method are similar to the *setMethodPolicy()* method but an extra parameter is required to specify the parameter to which the policy applies.
- *Class policy rules*, set using the *setClassPolicy()* method, are associated with classes rather than methods and indicate how instances of particular classes should be transmitted. For example, a class policy rule might specify that all instances of a particular class should be passed by-value. Class policy rules are applied based on the actual classes of the transmitted objects, rather than the classes specified in the method signature, which may be super-classes of the arguments. Class policy rules do not take a depth parameter since the object classes they reference may have a class policy associated with them.
- The *setFieldToCache()* method is used to indicate that a particular field in a particular class should be cached in remote references to instances of that class. The parameters to this method comprise the identity of the field to cache and the identities of the accessor methods of that field, which are also cached. Calls to these accessor methods are not propagated across the network but instead access the locally stored copy of the field.

- The *setMethodToCache()* method is used to indicate that a particular method in a particular class should be cached in remote references to instances of that class. Any calls to cached methods will be performed locally with respect to the caller.

An application programmer may specify or change policy rules at run-time, thus allowing for dynamic adaptation of the application. To specify policy rules statically, a library class programmer can specify the policy rules in the class initialization code. The policy manager can also be configured to read and write policy rules stored in XML files, allowing the programmer to specify policies completely independently of the application source, as well as library class source.

Clearly, there is scope for contention between policy rules. For example, if an instance of class *X* is passed as a parameter to method *m()* then a class policy rule may indicate that instances of *X* are passed-by-value while a method policy rule simultaneously indicates that parameters to method *m()* are passed-by-reference. As shown in Figure 8, each rule has a particular priority.

When contention occurs, the highest priority rule that applies is chosen over all others. An order of precedence is also imposed on policy rules based on their types to allow the framework to choose between rules of different types with the same priority. This approach to rule priority and precedence ensures that the temporal order in which rules are specified is not relevant, which is important given that policy rules may be defined dynamically in arbitrary application classes at any time during execution.

## 6.2 Revisiting the Example

In our peer-to-peer example introduced earlier, a *Message* might be transmitted by-value to an end-point using the *route* method on a *P2PNode*. However, if some of these objects are very large, the client programmer may wish to transmit them by-reference. Figure 9 shows how the *deliver()* method from Figure 6 may be modified to use the *transmission policy manager* to send those *Message* objects which exceed some maximum size by-reference, and smaller *Message* objects by-value.

```
public void deliver(Key dest,
  Message msg) throws Exception {
  IP2PNode node = (IP2PNode)
    remoteRRT.getRemoteReference("P2P");
  TransmissionPolicyManager tpm =

localRRT.getTransmissionPolicyManager();
  if (message.getSize() > MAX) {
    tpm.setClassPolicy(Message.class,
      BY_REF, 0);
  } else {
    tpm.setClassPolicy(Message.class,
      BY_VALUE, 0);
  }
  node.route(destination, message);
}
```

**Figure 9: The modified *deliver()* method.**

In the P2P application, instances of the *Key* class are immutable. Figure 10 illustrates the code necessary to instruct the transmission policy manager to make proxies to instances of class *P2PNode* cache the field *key*. This code fragment also specifies a class policy rule indicating that instances of class *Key* should be passed by-value. On the client-side, the call to *getRemoteReference()* will yield a proxy of the remote *P2PNode* object which can be cast to the remote type *IP2PNode*. A client holding such a proxy can access the *key* value of the remote *P2PNode* without incurring the cost of a remote call.

```
Method getKeyMethod = P2PNode.class.
  getDeclaredMethod("getKey", null);
Field keyField = P2PNode.class.
  getDeclaredField("key");
tpm.setFieldToCache(keyField,
  getKeyMethod, null);
tpm.setClassPolicy(Key.class, BY_VALUE,
0);
```

**Figure 10: Defining a smart proxy for *P2PNode* objects.**

# 7. IMPLEMENTATION ISSUES

The exposure of an object requires several steps. Firstly a *service adaptor* of the appropriate class is created. A *service adaptor* is the boundary between the application object and the Web Services infrastructure. There is one *service adaptor* class associated with each application class and one instance of a *service adaptor* class is created and associated with each exposed object. Thus there is a one-to-one correspondence between *service adaptors* and services. A *service map* maps from names and GUIDs to the *service adaptors* associated with the particular services. The RRT provides a generic *service adaptor* implementation that employs reflective techniques to invoke methods on exposed objects. Alternately, the RRT can automatically generate customized *service adaptor* classes which allow the RRT to perform method calls on them without using reflection. *Service adaptor* generation incurs a one time cost and obviates the need for reflection during normal execution. Generated code is cached in the RRT for the duration of the JVM lifetime but the RRT can be configured to cache generated code across multiple runs of the distributed application.

## 7.1 Serialisation

During the object marshalling phase of a remote method call, the RRT will determine which object transmission semantics to employ. If pass-by-value semantics have been chosen, the RRT will serialize the closures of the arguments. A generic serializer that can handle both primitive SOAP types, such as *ints* and *strings*, and complex types is provided. The RRT can be configured to automatically generate per-class serializers that are tuned to serialize instances of a particular application class.

Support for the transmission of arbitrary types is provided through an extension to Web Services semantics, which is incompatible with standard Web Services. The RRT employs the extended semantics when both client and server are RRT-based to allow full support for the transmission of sub-types. When the RRT is used in conjunction with conventional Web Services technology, standard Web Services semantics are adopted. The RRT determines whether to employ extended semantics on a per-call basis.

## 7.2 Implementing Remote References

The RRT implements remote references using remote identifiers, called *RAFDA Interoperable Object References* (*RafdaIORs*), and proxy objects. A *RafdaIOR* uniquely identifies an exposed service in the distributed system and consists of:

- The socket address of the RRT instance exposing the object. When remote method calls are performed on the object, this address determines the RRT instance to which the remote method call requests are sent.
- A string representation of a 160-bit Universally Unique Identifier (UUID) that identifies the Web Service associated with the exposed object.
- An instance of *java.lang.Class* capturing the remote type associated with the object, which was specified at exposure time. This remote type is used client-side during proxy generation and indicates which methods provided by the object's class will be remotely accessible.
- An instance of *java.lang.Class* representing the class of the exposed object. This is identified as the *real class* to differentiate it from the object's remote type. This class is used during proxy generation.
- A list of the fields to be cached in any proxy objects associated with the exposed object, which is used during proxy generation.
- A list of the methods to be cached in any proxy objects associated with the exposed object, which is also used during proxy generation.
- The current values of any cached fields.

To pass objects by-reference, the RRT serializes the associated *RafdaIORs* by-value. On deserialization, the client-side RRT uses the *RafdaIORs* to create and initialize appropriate proxy objects. Proxies, like service adaptors and serializers, are automatically generated as required by the RRT. From the client's perspective, the proxy class is the same type as the remote type specified in the *RafdaIOR*. For every method in the remote type, the proxy implements an associated method with the same signature, which calls into the RRT to make a remote call to the exposed object on behalf of the client.

Application objects cannot make use of *RafdaIORs* directly; they can only use references to other application objects or correctly typed proxy objects that have been initialized with the *RafdaIORs*. Therefore, when *RafdaIORs* are received by RRTs during remote method calls, the RRTs convert them into references that the application can use.

Initially, the RRT determines whether the referenced object exists in the local address space and if it does then a direct reference to the object is passed to the application. If not, the RRT determines whether a proxy to the referenced object has already been instantiated in the local address-space and, if the proxy exists then a reference to it is passed into the application. If a proxy does not already exist, then an instance of the associated proxy class is instantiated, automatically generating the class if necessary. This approach avoids the unnecessary use of remote references that loop-back into the same address spaces or the instantiation of more proxies than necessary.

## 7.3 Smart Proxies

All RRT proxy objects are *smart proxies*, meaning that they are capable of caching some of the exposed objects' fields or code. *RafdaIORs* contain smart proxy information indicating which fields and methods should be cached in the proxy and from this, an appropriate proxy class can be generated. The proxy class inherits the cached fields and methods from the remote type and the cached fields' *get()* and *set()* methods are modified to access the fields locally rather than invoke the equivalent methods on the exposed object. Non-cached methods are overridden with proxy versions while cached methods are not overridden, leaving the original functionality in place. A new proxy class is generated for each combination of cached fields and methods in use within the distributed application.

Immediately before a *RafdaIOR* is serialized, the RRT records the current values of the cached fields in it and they are serialized as part of the *RafdaIOR*. On deserialization, the cached fields in the proxy object are initialized automatically.

The RRT does not provide any form of automatic coherency control and so the programmer has responsibility for ensuring that application semantics remain as expected. Caching is particularly useful when object fields are known to be immutable.

## 7.4 Automatic Exposure

The RRT can export references to objects that have not been exposed to remote access, for example, as return values or in the closure of returned objects. The RRT performs automatic exposure of any such referenced objects on demand. By default, the RRT exposes objects using their own classes as remote types, with automatically generated service names. However, the concept of remote types stems from the fact that it is not always desirable to expose all methods of a given object to remote access. Programmers can therefore associate particular remote types with particular application classes using the *associateClassWithRemoteType()* method provided by the *IRafdaRunTime* interface.

## 7.5 Remote Method Call Cost

The cost of remote method calls in the RRT prototype was compared with the equivalent calls using other middleware systems. A test application was created then distributed using multiple different middleware technologies.

Tests were run on a two machine network. The first machine, designated the *"server"*, was used to execute the server-side applications that exposed objects to remote access. It contained a 2.7GHz Pentium 4 with 512MB RAM. The second machine, designated the *"client"*, was used to execute the client-side applications that performed the remote calls. It contained a 1.2GHz Pentium 3 with 256MB RAM. The machines were connected using an isolated 100Mb/s Ethernet. Since the .NET framework executes only under the Windows operating system, all tests on both machines were run under Windows XP Service Pack 2, fully patched, with only default services running.

The first test evaluates the cost of a remote method call to a method that takes no arguments, performs no computation and returns no results. This test determines the lower bound of call cost, since there are no arguments or return values to pass, meaning no marshalling is performed. The clock resolution provided by the test machines is 10ms, which is considerably greater than the average method call time. Therefore the test application performs 100 batches of 4000 method calls using each middleware system, resulting in a total run-time of between two and twenty minutes wall clock time. The system clock is used to measure the time taken to perform each of the 100 batches of method calls. Apache Axis received special treatment as it runs around an order of magnitude slower than all other systems. Each batch performs only 400 method calls, rather than 4000, in order to achieve reasonable total test execution time.

The second test was run under the same conditions as the first test but introduces arguments that require serialization. The method called by this test application takes ten arguments, all of which are passed by-value. The arguments are all instances of the same complex type, which contains a 10 character string, a 25 character string and an integer. In all tests the arguments are initialized identically. Table 1 shows the average time in milliseconds for a remote method call in each test.

**Table 1: Time in milliseconds for a remote method call.**

| Middleware | Without Serialization | With Serialization |
|---|---|---|
| *Java RMI (J2SE 1.5)* | 0.26 | 0.43 |
| *Microsoft .NET (C# using TCP channel)* | 0.44 | 0.86 |
| *CORBA (J2SE 1.5 ORB)* | 0.87 | 1.41 |
| *RRT* | 2.10 | 2.63 |
| *Microsoft .NET (C# using SOAP channel)* | 2.94 | 5.07 |
| *Apache Axis (1.2 final)* | 12.60 | 20.88 |

A clear difference can be seen between the middleware systems that use XML-based SOAP as their transport protocol (the RRT, Apache Axis and the .NET framework employing SOAP channels) and those that use binary protocols (Java RMI, CORBA and the .NET framework employing TCP channels). The RRT outperforms both its SOAP-based counterparts; the application employing the RRT ran in around 75% of the time taken by the equivalent .NET application and around 15% of the time taken by the application employing Apache Axis. When serializing a large number of arguments, the RRT is again the quickest of the SOAP-based systems. During this test, the RRT used cached per-class serializers in order to optimize the serialization process, giving it a large advantage over the other systems, which do not generate such serializers.

The applications using Java RMI, CORBA and TCP-based .NET all executed two to five times as quickly as the RRT. It should be noted that there are many implementations of the CORBA specification and that the one tested is that supplied with the J2SDK 5.0. It is reasonable to suggest that commercial ORBs may be better tuned for performance than this implementation and that the call time could be reduced more in line with the other systems that employ binary protocols. While the middleware systems that employ binary protocols outperform the RRT, the binary approach has disadvantages in that it does not provide the meta-data and opportunities for validation that XML does. SOAP can be considered the safer approach as the data is self-describing and less prone to problems with type safety [4].

SOAP-based systems offer a high degree of interoperability and a transport protocol with multiple advantages over binary approaches, as discussed above. Of the SOAP systems tested, the RRT prototype performed best, indicating that the advantages provided by the RRT's approach to application creation need not come at the cost of degraded performance.

## 7.6 Implementation of the Transmission Policy Framework

The policy framework is implemented using six associative stores, one for each rule type. Each associative store records argument policy rules and maps from keys to prioritized lists of policy rules. The keys are deterministically generated from the identity of the class and method being called and the argument numbers (where appropriate). To determine if an argument policy exists, the policy manager looks up the associative stores in order and if a mapping from the specified key exists, then the dominant argument policy rule is used. This approach is both simple and efficient.

The policy framework must be queried and the policy rules evaluated each time objects are marshalled, affecting remote method call cost. This cost is heavily dependent on the particular policy rules that are associated with the object to be marshalled. The transmission policy framework is an integral part of the RRT and so cannot be switched off under normal circumstances. To determine the cost of transmission policy evaluation, a special build of the RRT that employed only pass-by-reference semantics was created.

A test application that performed multiple calls to a remote method was created. This method took one argument and returned one return value, both by-reference. The test application was run using the specially built RRT with the transmission policy framework removed and again using the full RRT. In the former case, the special RRT was hard-coded to pass objects by-reference, and in the latter case, the transmission policy consisted of a method policy rule and a return policy rule stating that pass-by-reference semantics should be employed. The parameter passing semantics were therefore the same for each run of the application.

The cost of a remote call when the policy evaluation phase was performed was around 2% to 3% greater than the cost of a remote call without the evaluation phase. The introduction of additional arguments has no effect on the proportionate cost of the policy evaluation phase as there is a one-to-one correspondence between the number of objects marshalled and the number of transmission policy evaluations performed. The cost of dynamically evaluating policy is subsumed by the cost of marshalling and serialising the objects for remote method call. It is believed that the benefits gained outweigh the expense.

# 8. CONCLUSIONS

The RAFDA Run-Time (RRT) is a middleware designed to improve the software engineering process for implementers of new distributed systems and monitoring/management infrastructures aimed at existing applications. The work described in this paper has identified a number of key limitations exhibited by standard middleware systems and had shown how the mechanisms provided by the RRT addresses each of these limitations.

Middleware systems typically require the programmer to decide at application design time which classes will support remote access and to follow a number of steps in order to create the remotely accessible classes. The programmer must decide the interfaces between distribution boundaries statically then determine which classes will implement these interfaces and thus be remotely accessible. This hard-coding of the distribution boundaries requires that the application programmer know if instances of a class will be remotely accessed before implementing that class.

Using the RRT, programmers can adopt a new methodology when developing and deploying distributed Java applications [23]. Application logic can be designed and implemented completely independently of distribution concerns, easing the development task and giving considerable flexibility to alter distribution decisions late in the development cycle.

The RRT permits instances of arbitrary classes within an application to be exposed for remote access. This is achieved through the dynamic exposure of a standard Web Service for the exposed object and the implementation of a mapping from remote calls on the Web Service to method calls on the exposed object. The RRT introduces pass-by-reference semantics to standard Web Services allowing methods on exposed objects to be called remotely.

In contrast to conventional middleware systems, in order to expose an instance of a class using the RRT, it is not necessary that the class implement any special interfaces or extend any special classes. Objects can be exposed to remote access using any interface with which they are structurally compliant. Thus the application programmer can implement the classes providing core application functionality without regard for the remote accessibility of the instances of those classes. Decisions about the remote accessibility of a particular object can be delayed until much later in the design cycle, even until run-time. Monitoring and management infrastructure that views and controls application state from another address space can be created without modification, or even access, to the application's original source code.

Another limitation of existing middleware systems is that the parameter passing semantics is tightly bound to the distribution of the application and thus changes to the distribution of an application may potentially alter the application semantics. The RRT addresses this limitation by providing a framework for the static and dynamic specification of object transmission policy. Using this framework the application programmer can employ the most advantageous object transmission policy for the particular circumstances. This increases flexibility and allows the programmer to control the application semantics. By specifying object transmission policy independently of class implementation, the roles of library class programmer and application programmer are separated. Library implementers need make fewer assumptions about the ways in which their classes will be used while application programmers can use class instances in the most appropriate way, as dictated by the particular situation. Before making a method call the application programmer can configure the transmission policy for the individual method parameters.

The transmission policy framework also supports the specification of smart proxies which increase the flexibility of exposed object without imposing implementation constraints on the programmer. This mechanism allows arbitrary field values of an object to be cached in the same address space as a remote reference (proxy) to that object. Thus a call to an accessor method on the proxy yields the cached field value without the execution of a network call.

The RRT employs dynamic code generation and compilation techniques to create the ancillary code necessary to allow dynamic object exposure. It is capable of marshalling instances of any class either by-reference or by-value and complete control over this is given to the programmer in order to separate parameter passing semantics completely from application distribution.

The RRT provides significant advantages to programmers of distributed applications, when compared to industry standard middleware systems, simplifying the software engineering process, decreasing the opportunity for errors in distribution code and increasing code reuse through better flexibility.

The RRT has been used in the construction of a P2P routing network in which the application code can be run in both a fully distributed environment and in a centralised simulation environment without modification.

The RAFDA system can be downloaded from *http://rafda.cs.st-and.ac.uk/*.